\documentclass[10pt,final,journal,a4paper,twoside,twocolumn,romanappendices]{IEEEtran}
\IEEEoverridecommandlockouts
\usepackage[dvips]{psfrag}
\usepackage[dvips]{epsfig}
\usepackage{epic,color}
\usepackage[cmex10]{amsmath}
\usepackage{amsfonts}
\usepackage{amssymb}
\usepackage{amsmath}
\usepackage{bm}
\usepackage{yfonts}
\usepackage{array}
\usepackage{bigstrut}
\usepackage{multirow}
\usepackage{dsfont}
\usepackage{cite}
\usepackage{fixltx2e}
\usepackage{accents}
\usepackage{mathtools}
\usepackage[normalem]{ulem}
\usepackage{enumerate}
\usepackage{stfloats}
\usepackage[T1]{fontenc}
\usepackage{float}
\usepackage{epstopdf}
\usepackage{mathtools}
\usepackage{algorithm}
\usepackage{algorithmicx}
\usepackage{algpseudocode}
\usepackage{algorithm,algpseudocode,float}
\usepackage{lipsum}
\usepackage{soul,xcolor}
\usepackage{xparse}
\usepackage{acronym}
\usepackage{soul}

\NewDocumentCommand{\ceil}{s O{} m}{%
  \IfBooleanTF{#1} 
    {\left\lceil#3\right\rceil} 
    {#2\lceil#3#2\rceil} 
}
\NewDocumentCommand{\floor}{s O{} m}{%
  \IfBooleanTF{#1} 
    {\left\lfloor#3\right\rfloor} 
    {#2\lfloor#3#2\rfloor} 
}

\newtheorem{proposition}{Proposition}
\newcommand{\txt}[1]{\texttt{#1}}
\newcommand{\mbf}[1]{\mathbf{#1}}
\newcommand{\ub}[1]{\underbrace{#1}}
\newcommand{\norm}[1]{\left\lVert#1\right\rVert}

\algrenewcommand\algorithmicrequire{\textbf{Input:}}
\algrenewcommand\algorithmicensure{\textbf{Output:}}

\newlength{\algorithmwidth}
\AtBeginDocument{\setlength{\algorithmwidth}{1.2\textwidth}}

\acrodef{5G}{fifth generation}
\acrodef{HetNet}{heterogeneous network}
\acrodef{UE}{user equipment}
\acrodef{HD}{half-duplex}
\acrodef{FD}{full-duplex}
\acrodef{BS}{base station}
\acrodef{MBS}{macro base station}
\acrodef{SBS}{small cell base station}
\acrodef{DL}{downlink}
\acrodef{UL}{uplink}
\acrodef{CSI}{channel state information}
\acrodef{MIMO}{multiple-input multiple-output}
\acrodef{SINR}{signal-to-interference plus noise ratio}
\acrodef{AWGN}{additive white Gaussian noise}
\acrodef{MMSE}{minimum mean square error}
\acrodef{SIC}{successive interference cancellation}
\acrodef{SI}{self-interference}
\acrodef{CCI}{co-channel interference}
\acrodef{MUI}{multiuser interference}
\acrodef{SOC}{second-order cone}
\acrodef{SOCP}{second-order cone program}
\acrodef{RAOFDS}{resource allocation optimization for full-duplex small cell}
\acrodef{SPCA}{successive parametric convex approximation}
\acrodef{PSD}{positive semi-definite}
\acrodef{PPP}{Poisson point process}
\acrodef{DE}{decoding energy}
\acrodef{EH}{energy harvesting}
\acrodef{w.r.t.}{with respect to}
\acrodef{DD}{dual decomposition}
\acrodef{ADMM}{alternating direction method of multipliers}

\begin{document}

\title{Distributed Energy and Resource Management for Full-Duplex Dense Small Cells for 5G}
\author{\IEEEauthorblockN{Animesh Yadav,
Octavia A. Dobre and Nirwan Ansari\IEEEauthorrefmark{1}}\\
\IEEEauthorblockA{Faculty of Engineering and Applied Science, Memorial University, St. John's, NL, Canada\\ \IEEEauthorrefmark{1}Dept. of Electrical and Computer Engineering, New Jersey Institute of Technology, Newark, NJ, USA\\
Email: \{animeshy, odobre\}@mun.ca, nirwan.ansari@njit.edu}}



\maketitle
\thispagestyle{empty}
\begin{abstract}
We consider a multi-carrier and densely deployed small cell network, where small cells are powered by renewable energy source and operate in a full-duplex mode. We formulate an energy and traffic aware resource allocation optimization problem, where a joint design of the beamformers, power and sub-carrier allocation, and users scheduling is proposed. The problem minimizes the sum data buffer lengths of each user in the network by using the harvested energy. A practical uplink user rate-dependent decoding energy consumption is included in the total energy consumption at the small cell base stations. Hence, harvested energy is shared with both downlink and uplink users. Owing to the non-convexity of the problem, a faster convergence sub-optimal algorithm based on successive parametric convex approximation framework is proposed. The algorithm is implemented in a distributed fashion, by using the alternating direction method of multipliers, which offers not only the limited information exchange between the base stations, but also fast convergence. Numerical results advocate the redesigning of the resource allocation strategy when the energy at the base station is shared among the downlink and uplink transmissions.
\end{abstract}
%

\section{Introduction}
For the year 2020 and beyond, the \ac{5G} mobile communications technology has promised to provide a 1000-fold increase in data rate and enhanced user experience. Among the technologies that have the potential to achieve the \ac{5G} promises are the dense deployment of small cells \cite{Bhushan-Li-Malladi-Gilmore-Brenner-Damnjanovic-Sukhavasi-Patel-Geirhofer-mcom-15} and \ac{FD} communications. Small cells are energy- and cost-efficient \acp{BS} that bring the users closer to them, and thus, increase the network throughput and user experience. On the other hand, the \ac{FD} technique is rekindled to utilize the spectrum efficiently. The \ac{FD} communications essentially allows the simultaneous transmission and reception of signals on the same time-frequency resource, and thus, improves the spectral efficiency of the network. The benefit of using the two technologies simultaneously is evident, but with a few challenges.  

In a densely and arbitrarily deployed network scenario, the incumbent operators might face difficulties in powering the \acp{SBS} through the grid power source. Hence, alternately, they can install energy harvested device to each \ac{SBS} for harvesting the energy from nature \cite{Huang-Han-Ansari-CST-2015,he2014recursive}. This approach is not only environment-friendly by curbing the $\text{CO}_2$ emission, but also economical. Renewable energy can be freely harvested from nature using solar and wind sources. The amount and arrival of harvested energy are random by nature, thus sometimes leading to service interruption.  Hence, to reap the benefits of the freely available energy, the harvested energy must be used intelligently. With this objective, the communication system is designed with consideration of an intermittent source of energy \cite{Ahmed-Ikhlef-Ng-Schober-twcom-13,Arafa-Ulkus-jsac-2015, Yadav-Nguyen-Ajib-tcom-2016,he2016optimal,he2016non}.

Owing to the hardware incapability to handle \ac{SI}, the \ac{FD} technique, though conceptualized a long time ago, has not been used. Recently, efforts have been made to cancel \ac{SI} in both analog and digital domains jointly, e.g., \cite{Hong-others-CM-2014, Bharadia-Mcmilin-Katti-sigcomm-2013}, such that \ac{FD} communications become a  reality. However, these works advocate the applicability of \ac{FD} communications for short range, where the transmit power is low. Hence, the \acp{SBS} are the suitable candidates to operate in the \ac{FD} mode \cite{Nguyen-Tran-Pirinen-Latva-aho-twc-2014, Goyal-Liu-Panwar-tvt-2016}. Furthermore, since small cells have a range of operation of approximately 100 meters, the energy spent in decoding the received data is non-negligible \cite{Cui-Goldsmith-Bahai-jsac-04}. Hence, the \ac{SBS} has to share the available energy not only with the transmitter but also with the receiver operations.

To reap the benefits of simultaneously using the \ac{EH} \ac{SBS} and \ac{FD} communications, engineers face a few challenges: i) mitigation of interference surge due to \ac{FD} communications and ii) efficient sharing of harvested energy among the transceiver operations, such as transmitting energy and rate-dependent decoding energy. At the network level, the interference intensity is high when compared with single cell scenario, due to both intra- and inter-cell interference. A few works \cite{Goyal-Liu-Panwar-tvt-2016, Chen-other-jsac-2016} studied the increase of inter-cell  interference when the \ac{BS} in each cell is deployed with an \ac{FD} transceiver. Furthermore, the energy availability at the \ac{SBS} is random and needs to be shared among the transmitter and receiver operations optimally. Hence, recent works \cite{Rubio-Pascual-Iserte-twc-2014, Arafa-Ulkus-jsac-2015, Yadav-Nguyen-Ajib-tcom-2016, Yadav-Dobre-Ansari-access-2017} accounted for the received data rate-dependent \ac{DE} in their problems for a more realistic formulation. \ac{DE} is required to process the received data that are protected by some outer code, such as turbo or low-density parity check codes. 

In this paper, we consider the realistic communication scenario, where densely deployed \ac{EH} \acp{SBS}, operating in the \ac{FD} mode, serve \ac{HD} \ac{DL} and \ac{UL} \acp{UE}. 
The practical rate-dependent decoding power usage is included in the total power consumption at the \acp{SBS}\cite{Rubio-Pascual-Iserte-twc-2014}, and hence, the achieveable rates obtained by \ac{UL} \acp{UE} are not only dependent on the \acp{UE} power, but also on that of the \ac{SBS}. As a consequence, the solutions obtained in all previous works are not anymore applicable. Furthermore, to avoid the excessive resource allocations, also aligned with operators interest, we assume another realistic assumption of non-uniform wireless traffic, i.e., each \ac{UE} has different amount of data in its buffer to be transmitted. Thus, with the goal of efficiently managing the network resources in an excessive surge of interference due to \ac{FD} communications and under the random energy availability, we formulate the problem of jointly designing the transmit beamformer, power and sub-carrier allocation, and \acp{UE} scheduling. Moreover, distributively solving the optimization problem is of utmost important, especially for a dense network, which requires huge information exchange among the \acp{BS}. The centralized and dual decomposition based distributed algorithms to solve the problem are discussed in \cite{Yadav-Dobre-Ansari-access-2017}. Since the dual decomposition approach suffers from slow convergence, we propose to use a fast convergent \ac{ADMM} \cite{Boyd-others-FTML-2011} approach. In this approach, we decompose the problem into \ac{SBS} sub-problems by introducing the set of global variables that link the same variables of the coupled \acp{SBS}, i.e., the consensus equality constraints.

The rest of the paper is organized as follows. Section II introduces the system model and formulates the optimization problem. Section III develops an algorithm based on the \ac{ADMM} framework to distributively solve the optimization problem. Section IV presents numerical results and discussions. Finally, conclusion of the paper is given in Section V.
%

\section{System Model and Problem Formulation}
\subsection{System Model}
A multi-carrier multi-cell network consisting of $B$ \ac{EH} \ac{FD} \acp{SBS} serving \ac{HD} \acp{UE} is considered in \cite{Yadav-Dobre-Ansari-access-2017}. Each \ac{SBS} is installed with a rechargeable battery and an \ac{EH} device, which are used to store and collect the harvested energy, respectively. The \acp{SBS} are equipped with $M_T + M_R$ antennas, of which $M_T$ antennas are used to transmit data on the \ac{DL} channel and $M_R$ antennas are used to receive data on the \ac{UL} channel. Each base station $b$ belongs to a set denoted by $\mathcal{B}=\{1,\ldots,B\}$.  The sets of all \ac{DL} and \ac{UL} \acp{UE} are denoted by $\mathcal{D} = \{1, \ldots, K_{\text{D}}\}$ and  $\mathcal{U} = \{1, \ldots, K_{\text{U}}\}$, respectively. We assume that data for the \ac{DL} \ac{UE} $i$ are transmitted only from one \ac{SBS}, and are denoted by $b_i\in \mathcal{B}$. Similarly, the data of \ac{UL} \ac{UE} $j$ are processed by only one \ac{SBS}, and are denoted by $b_j\in \mathcal{B}$. The sets of all \ac{DL} and \ac{UL} \acp{UE} associated with \ac{SBS} $b$ are denoted by $\mathcal{D}_b \in \mathcal{D}$ and $\mathcal{U}_b \in \mathcal{U}$, respectively. The \acp{SBS} send and receive data simultaneously to $K_{\text{D}}$ \acp{UE} on the DL channels and from $K_{\text{U}}$ \acp{UE} on the UL channels, respectively. We further assume that the \ac{MBS} is serving the \acp{UE} on the \ac{UL} channels. A total of $N$ equal bandwidth sub-channels belonging to the set $\mathcal{N} = \{1. \ldots, N\}$ are available in the system. 

The received signal over sub-channel $n$ at \ac{DL} \ac{UE} $i$ is given by
\begin{IEEEeqnarray}{lll}
y^{\txt{D}}_{i,n} &= \mbf{h}^H_{b_{i},i,n}\mbf{u}_{i,n}s^{\txt{D}}_{i,n} + \ub{\sum_{k\neq i}^{K_{\txt{D}}}\mbf{h}^H_{b_{k},i,n}\mbf{u}_{k,n}s^{\txt{D}}_{k,n}}_{\text{MUI + CCI due to all DL UEs}} \IEEEnonumber
\end{IEEEeqnarray}
\begin{IEEEeqnarray}{lll}  
& + \ub{\sum_{j=1}^{K_{\txt{U}}}g_{j,i,n}\sqrt{p_{j,n}}s^{\txt{U}}_{j,n}}_{\text{CCI due to all UL UEs}} + n^{\txt{D}}_{i,n},\IEEEyesnumber
 \label{eq:Rx_signal_DL} 
\end{IEEEeqnarray}  
where $\mbf{u}_{i,n}$ and $p_{j,n}$ are the beamforming vector and power coefficient corresponding to the \ac{DL} and \ac{UL} \acp{UE} $i$ and $j$, respectively, on the $n$th sub-channel. $\mbf{h}_{b_{i},i,n} \in \mathbb{C}^{M_T\times 1}$ is the channel vector from \ac{SBS}  $b_i$ to \ac{DL} \ac{UE} $i$ and $g_{j,i,n}$ is the complex channel coefficient from \ac{UL} \ac{UE} $j$ to \ac{DL} \ac{UE} $i$ on the sub-channel $n$. $s^{\txt{D}}_{i,n}$ and $s^{\txt{U}}_{j,n}$ are data symbols corresponding to the \ac{DL} and \ac{UL} \acp{UE}, respectively, each with unit average energy, i.e., $\mathbb{E}\{|s^{\txt{D}}_{i,n}|^2\}=1$. $\mathbb{E}\{\cdot\}$ denotes the expectation operator. The term $n^{\txt{D}}_{i,n} \sim \mathcal{CN}(0,\sigma_n^2)$ is the \ac{AWGN}. In \eqref{eq:Rx_signal_DL}, the first and second terms on the right-hand side represent the intended signal and the sum of intra-cell \ac{MUI} and inter-cell \ac{CCI} due to all \ac{DL} transmissions, respectively. The third term represents the \ac{CCI} due to all \ac{UL} transmissions. The received \ac{SINR} of \ac{DL} \ac{UE} $i$ over sub-channel $n$ can be written as 
\begin{align} \label{eq:SINR_DL}
&\gamma^{\txt{D}}_{i,n} =  \cfrac{\mbf{h}^H_{b_{i},i,n}\mbf{U}_{i,n}\mbf{h}_{b_{i},i,n}}{\sigma_n^2 + \sum_{k\neq i}^{K_{\txt{D}}}\mbf{h}^H_{b_{k},i,n}\mbf{U}_{k,n}\mbf{h}^H_{b_{k},i,n} +  \sum_{j=1}^{K_{\txt{U}}}p_{j,n}|g_{j,i,n}|^2},
\end{align}
where $\mbf{U}_{i,n} = \mbf{u}_{i,n}\mbf{u}^H_{i,n}$ is a \ac{PSD} matrix.

Next, for the \ac{UL} transmission, the received signal vector of \ac{UE} $j$ over sub-channel $n$ at \ac{BS} $b_j$ is given by
\begin{IEEEeqnarray}{lll}
\mbf{y}^{\txt{U}}_{j,n} &= \displaystyle \mbf{h}_{b_j,j,n}\sqrt{p_{j,n}}s^{\txt{U}}_{j,n} + \sum_{l\neq j}^{K_{\txt{U}}}\mbf{h}_{b_j,l,n}\sqrt{p_{l,n}}s^{\txt{U}}_{l,n} \IEEEnonumber \\ &+ \ub{\sum_{i=1}^{K_{\txt{D}}}\mbf{H}_{b_j,b_i,n}\mbf{u}_{i,n}s^{\txt{D}}_{i,n}}_{\text{SI + CCI from all DL UEs}} + \mbf{n}^{\txt{U}}_{j,n},
\label{eq:Rx_signal_UL} 
\end{IEEEeqnarray}  
where $\mbf{h}_{b_{j},j,n}\in \mathbb{C}^{M_R \times 1}$ is the channel vector from \ac{UL} \ac{UE} $j$ to \ac{SBS} $b_j$ and $\mbf{n}^{\txt{U}}_{j,n}\sim \mathcal{CN}(0,\sigma_n^2\mbf{I}_{M_R})$ is the \ac{AWGN} noise vector. In \eqref{eq:Rx_signal_UL}, the first right-hand side term is the intended signal. The second right-hand side term represents the intra-cell multiple access interference and inter-cell \ac{CCI} due to all \ac{UL} transmissions. The third term  represents the total \ac{CCI} due to inter-cell \ac{DL} transmissions including \ac{SI}, where $\mbf{H}_{b_j,b_i,n}$ is the channel matrix from \ac{SBS} $b_j$ to \ac{SBS} $b_i$.
In order to recover each \ac{UL} \ac{UE} data, we treat the \ac{SI} and \ac{CCI} as background noise and apply the \ac{MMSE} successive interference cancellation receiver. Then, the received \ac{SINR} of \ac{UL} \ac{UE} $j$ over sub-channel $n$ is given by
\begin{IEEEeqnarray}{lll}
\gamma^{\txt{U}}_{j,n} &= \displaystyle p_{j,n}\mbf{h}^H_{b_j,j,n}\bigg(\sigma_n^2\mbf{I}_{M_R} + \sum_{l> j}^{K_{\txt{U}}}p_{l,n}\mbf{h}_{b_j,l,n}\mbf{h}^H_{b_j,l,n} \IEEEnonumber \\
&+ \displaystyle\sum_{i=1}^{K_{\txt{D}}}\mbf{H}_{b_j,b_i,n}\mbf{U}_{i,n}\mbf{H}^H_{b_j,b_i,n}\bigg)^{-1}\mbf{h}_{b_{j},j,n}.
\label{eq:SINR_UL} 
\end{IEEEeqnarray}  

We denote the number of backlogged bits waiting in the data buffer of \ac{DL} \ac{UE} $i$ at the given scheduling instant by $Q^{\txt{D}}_i$. At that instant, the reduction in backlogged bits achieved by the $i$th \ac{UE} is expressed as
\begin{eqnarray}
q^{\txt{D}}_{i} = Q^{\txt{D}}_i - \sum_{n=1}^{N}\log_2(1+\gamma^{\txt{D}}_{i,n}),
\label{eq:Q_DL} 
\end{eqnarray}   
where the second right-hand side term is the transmission rate achieved by \ac{DL} \ac{UE} $i$.
Similarly, on the \ac{UL} channel, the reduction in backlogged bits achieved by the \ac{UL} \ac{UE} $j$ is given by
\begin{eqnarray}
q^{\txt{U}}_{j} = Q^{\txt{U}}_j - \sum_{n=1}^{N}\log_2(1+\gamma^{\txt{U}}_{j,n}),
\label{eq:Q_UL} 
\end{eqnarray}
where $Q^{\txt{U}}_{j}$ denotes the number of backlogged bits corresponding to \ac{UL} \ac{UE} $j$ and the second right-hand side term represents the number of transmitted bits by \ac{UL} \ac{UE} $j$. 

\subsection{Energy Arrival and Usage Model}
We consider a generic renewable energy source, at each \ac{SBS}, such that the analysis presented in the sequel is valid for any energy arrival process. Let $B_{\text{max}}$ denote the maximum size of the rechargeable battery, which is used to store the sum of the energy harvested, i.e., $P_{b,\text{H}}$ and the leftover energy $P_{b,\text{B}}$ over the current and from the previous scheduling periods, respectively. Furthermore, at the beginning of the next scheduling period, the exact amount of energy available in the battery is known at the \ac{SBS}. Hence, for a given scheduling period, the energy available at the \ac{SBS} $b$ is given as $TP_{b}=\min\{B_{\text{max}}, TP_{b,\text{H}}+TP_{b,\text{B}} \}$, where $T$ is the length of a scheduling period in seconds and the $\min(\cdot, \cdot)$ operator ensures the constraint on the maximum battery size.

In short-distance communications, the energies consumed in the circuit and decoding become comparable or even dominate the actual transmit power \cite{Cui-Goldsmith-Bahai-jsac-04}. Hence, it is important to include them into the total power consumption, especially when the energy comes from a renewable source. The total power consumption at an \ac{SBS} is expressed as:
\begin{eqnarray}
P_{\text{tot},b} = \sum_{n=1}^{N}\sum_{i\in \mathcal{D}_b}\text{tr}(\mbf{U}_{i,n})+P_{b}^{\text{cir}} + \sum_{n=1}^{N}\sum_{j \in \mathcal{U}_b}P_{j,n}^{\text{dec}}(R_{j,n}),
\label{eq:power_consumption_SC} 
\end{eqnarray} 
where $P_{b}^{\text{cir}} = M_TP_{\text{rf}} + P_{\text{st}}$ is the total circuit power consumption, in which $P_{\text{rf}}$ and $P_{\text{st}}$ correspond to the active radio frequency blocks, and to the cooling and power supply, respectively. $P_{j,n}^{\text{dec}}$ is the power consumption for decoding \ac{UL} \ac{UE} $j$ in sub-carrier $n$, where $R_{j,n}=\log_2(1+\gamma^{\txt{U}}_{j,n})$ is the achievable rate of the \ac{UE}. Note that the decoding power consumption is a function of the data rate of the \ac{UE}: for example, for an UL \ac{UE} $j$, $P_{j,n}^{\text{dec}}(R_{j,n}) = \alpha_jR_{j,n}$ where $\alpha_j$ models the decoder efficiency, being decoder specific \cite{Cui-Goldsmith-Bahai-jsac-04, Rubio-Pascual-Iserte-twc-2014}.

\subsection{Optimization Problem Formulation}
In this work, we are interested in reducing the total number of backlogged bits in the network by minimizing the $\ell_2$-norm of the deviation metrics given in \eqref{eq:Q_DL} and \eqref{eq:Q_UL} \cite{Yadav-Dobre-Ansari-access-2017}. The main reason for using the $\ell_2$-norm in the objective function is that it gives priority to the \ac{UE} with a large queued data in the buffer.

Now, by denoting $\mbf{U} = [\mbf{U}_1, \ldots, \mbf{U}_{B}]$, where $\mbf{U}_b = [\mbf{U}_{\mathcal{D}_b(1),1},\ldots, \mbf{U}_{\mathcal{D}_b(|\mathcal{D}_b|),N}]$\footnote{$\mathcal{A}(i)$ and $|\mathcal{A}|$ denote the $i$th element and cardinality of set $\mathcal{A}$, respectively.} and $\mbf{p} = [\mbf{p}_1, \ldots, \mbf{p}_B]$, where $\mbf{p}_b = [p_{\mathcal{U}_b(1),1},\ldots, p_{\mathcal{U}_b(|\mathcal{U}_b|),N}]$, the optimization problem to be solved at the beginning of each scheduling period is formulated as
\begin{IEEEeqnarray*}{lcl}\label{eq:Problem1}
&\displaystyle\underset{\begin{subarray}{c}\mbf{U},\mbf{p}\end{subarray}}{\text{min}}\quad &\norm{\mbf{q}_{\txt{D}}}_2 + \norm{\mbf{q}_{\txt{U}}}_2 \IEEEyesnumber
\IEEEyessubnumber* \label{eq:problem_orig}\\
&\text{s.t.}&\sum\limits_{n=1}^N\sum\limits_{i \in \mathcal{D}_b}\text{tr}(\mbf{U}_{i,n})   \leq P_{b,\text{max}} \hfill \qquad\qquad \forall b, \label{eq:SBS_power_constr}\\
&&  P_{\text{tot},b} \leq P_b \hfill \forall	b, \label{eq:SBS_EH_constr}\\
&&\sum\limits_{n=1}^{N}p_{j,n} \leq P_{u,\text{max}}\hfill\forall j\in \mathcal{U}, \label{eq:UE_power_constr}\\
&& \text{rank}(\mbf{U}_{i,n}) = 1 \hfill \forall i \in \mathcal{D}, \forall n, \label{eq:UE_rank_constr}\\
&& \mbf{U}_{i,n} \succeq 0 \hfill \forall i \in \mathcal{D},  \forall n, \label{eq:PSD_constr}\\
&&p_{j,n}\geq 0  \label{eq:power_positive_constr}\hfill\forall j\in \mathcal{U}, \forall n,
\end{IEEEeqnarray*} 
where $\mbf{q}_{\txt{D}}$ and $\mbf{q}_{\txt{U}}$ have the elements $q^{\txt{D}}_i$ and  $q^{\txt{U}}_{j}$, respectively. $P_{b,\text{max}}$ is the maximum total transmit power constraint on the \ac{DL} channel, and $P_{u,\text{max}}$ is the individual \ac{UE} transmit power constraint on the \ac{UL} channel. It is worth noting that \eqref{eq:Problem1}\footnote{Note that \eqref{eq:Problem1} represents equations \eqref{eq:problem_orig}-\eqref{eq:power_positive_constr}. A similar notation is employed throughout the paper.} implicitly solves the problem of sub-carrier allocation and \ac{UE} scheduling as well. Hence, the optimization problem jointly designs the beamformers, power and sub-carrier allocation and \ac{UE} scheduling. An \ac{UE} is scheduled whenever it is allocated a non-zero power on a sub-carrier; otherwise, it is not.

In \eqref{eq:Problem1}, the objective function \eqref{eq:problem_orig} ensures avoidance of the redundant resource allocation, which is limited by the data queue length of the \acp{UE}. Further, constraint \eqref{eq:SBS_power_constr} ensures that the maximum transmit power allowed by \ac{SBS} $b$ for the \ac{DL} transmission is limited by $P_{b,\text{max}}$. Constraint \eqref{eq:SBS_EH_constr}  ensures the available energy at the \ac{SBS} is drawn by both the transmitter and receiver operations, and the energy causality constraint. In general, it is difficult to solve the above optimization problem due to the rank-one constraint. Hence, we relax the rank-one constraint and express the relaxed problem as 
\begin{IEEEeqnarray*}{rCl}\label{eq:relax_Problem1}
&\displaystyle\underset{\begin{subarray}{c}\mbf{U},\mbf{p}\end{subarray}}{\text{minimize}}  & \{\norm{\mbf{q}_{\txt{D}}}_2 + \norm{\mbf{q}_{\txt{U}}}_2\mid\eqref{eq:SBS_power_constr}-\eqref{eq:UE_power_constr}, \eqref{eq:PSD_constr},  \eqref{eq:power_positive_constr}\}. \IEEEyesnumber
 \label{eq:relax_problem_orig}
\end{IEEEeqnarray*} 
Owing to the non-concave objective function and constraint \eqref{eq:SBS_EH_constr} in \eqref{eq:relax_Problem1}, we propose to solve it by using the \ac{SPCA} method \cite{Beck-Ben-Tal-Tetruashvili-jgo-2010}. In this method, \eqref{eq:relax_Problem1} is successively approximated to a convex problem as presented in Proposition~\ref{prop:Approx_prop}, to obtain progressively improved solution.
 
\begin{proposition}\label{prop:Approx_prop}
By introducing the auxiliary variables $\bm{\beta}_b$, $\mbf{t}_b$, $\mbf{x}_b$, and $\mbf{z}_b$ for all $b\in\{1,\ldots,B\}$, the convex approximate of \eqref{eq:relax_Problem1}, at the $r$th \ac{SPCA} iteration, is expressed as
\begin{IEEEeqnarray*}{lCl}\label{eq:approx_problem3}
&\displaystyle\underset{\Xi}{\text{min}} \, & \norm{\tilde{\mbf{q}}_{\txt{D}}}_2 + \norm{\tilde{\mbf{q}}_{\txt{U}}}_2 \IEEEyesnumber
\IEEEyessubnumber* \label{eq:approx_problem3_obj}\\
&\text{s.t.}& \mbf{h}^H_{b_{i},i,n}\mbf{U}_{i,n}\mbf{h}_{b_{i},i,n} \geq  F(z^{\txt{D}}_{i,n},\beta_{i,n},\xi^{[r]})\hfill \forall i\in \mathcal{D},\forall n,\qquad \label{eq:approx_problem3_constr1}\\
&& H(x_{j,n}, \mbf{p}_{{\mathcal{U}\setminus \{j\}}}, \mbf{U}, x_{j,n}^{[r]}, \mbf{p}_{{\mathcal{U}\setminus \{j\}}}^{[r]}, \mbf{U}^{[r]}) \leq  z^{\txt{U}}_{j,n} \IEEEnonumber \\
&& \hfill \forall j\in \mathcal{U},\forall n, \qquad \label{eq:approx_problem3_constr2} \\
&&\sum\limits_{n=1}^N\sum\limits_{i \in \mathcal{D}_b}\text{tr}(\mbf{U}_{i,n})  \leq P_{b,\text{max}} \hfill \forall b,\qquad \label{eq:approx_problem3_constr3}
\end{IEEEeqnarray*}
\begin{IEEEeqnarray*}{lCl}
&\phantom{\text{s.t.}}&P_b^{\text{cir}} + \sum\limits_{n=1}^{N}\sum\limits_{j\in \mathcal{U}_b}\alpha_jt^{\txt{U}}_{j,n}  + \sum\limits_{n=1}^N\sum\limits_{i \in \mathcal{D}_b}\text{tr}(\mbf{U}_{i,n}) \leq P_b  \hfill\forall	b,\qquad \IEEEyessubnumber*\label{eq:approx_problem3_constr4}\\
&& e^{t^{\txt{D}}_{i,n}} \leq  z^{\txt{D}}_{i,n}+1\hfill \forall i\in \mathcal{D},\forall n,\qquad \label{eq:approx_problem3_constr5}\\
&&\sigma_n^2 + \sum_{k\neq i}^{K_{\txt{D}}}\mbf{h}^H_{b_{k},i,n}\mbf{U}_{k,n}\mbf{h}_{b_{k},i,n} + \sum_{j=1}^{K_{\txt{U}}}p_{j,n}|g_{j,i,n}|^2 \leq  \beta_{i,n}\IEEEnonumber \\
&&\hfill \forall i\in \mathcal{D},\forall n, \qquad\label{eq:approx_problem3_constr6}\\
&& p_{j,n} \geq  x^2_{j,n} \hfill \forall j\in \mathcal{U},\forall n,\qquad \label{eq:approx_problem3_constr7} \\
&& e^{t^{\txt{U}}_{j,n}} \leq  z^{\txt{U}}_{j,n}+1 \hfill \forall j\in \mathcal{U},\forall n,\qquad \label{eq:approx_problem3_constr8}\\
&& \eqref{eq:UE_power_constr},\, \eqref{eq:PSD_constr},\,\eqref{eq:power_positive_constr},
\label{eq:approx_problem3_constr9}
\end{IEEEeqnarray*}
\end{proposition} 
where $F(z^{\txt{D}}_{i,n},\beta_{i,n},\xi^{[r]}) = \beta^2_{i,n}/(2\xi^{[r]}) + \xi^{[r]}(z^{\txt{D}}_{1,n})^2/2$ and $H(x_{j,n}, \mbf{p}_{{\mathcal{U}\setminus \{j\}}}, \mbf{U}, x_{j,n}^{[r]}, \mbf{p}_{{\mathcal{U}\setminus \{j\}}}^{[r]}, \mbf{U}^{[r]})$ is a convex approximate of function $x^2_{j,n}\mbf{h}^{H}_{b_j,j,n}\mbf{X}^{-1}_{j,n}\mbf{h}_{b_j,j,n}$ at the $r$th iterate, where $\mbf{X}_{j,n} \triangleq \sigma^2_n\mbf{I}_{M_R} +\sum^{K_{\txt{U}}}_{l>j}p_{l,n}\mbf{h}_{b_j,l,n}\mbf{h}^H_{b_j,l,n} + \sum^{K_{\txt{D}}}_{i=1}\mbf{H}^H_{b_j,b_i,n}\mbf{W}_{i,n}\mbf{H}_{b_j,b_i,n}$. $\Xi=\{\Xi_1,\ldots,\Xi_{B}\}$ and $\Xi_b$ collects the variables corresponding to the \ac{BS} $b$, i.e., $\{\mbf{U}_b, \mbf{p}_b, \bm{\beta}_b, \mbf{t}_b, \mbf{x}_b,\mbf{z}_b\}$, where $\bm{\beta}_b=[\beta_{\mathcal{D}_b(1),1},\ldots,\beta_{\mathcal{D}_b(|\mathcal{D}_b|),N}]$, $\mbf{t}_b = [t^{\txt{D}}_{\mathcal{D}_b(1),1}, \ldots,t^{\txt{D}}_{\mathcal{D}_b(|\mathcal{D}_b|),N}, t^{\txt{U}}_{\mathcal{U}_b(1),1}\ldots,  t^{\txt{U}}_{\mathcal{U}_b(|\mathcal{U}_b|),N}]$, $\mbf{x}_b=[x_{\mathcal{U}_b(1),1},\ldots,x_{\mathcal{U}_b(|\mathcal{U}_b|),N}]$, and $\mbf{z}_b = [z^{\txt{D}}_{\mathcal{D}_b(1),1},\ldots,z^{\txt{D}}_{\mathcal{D}_b(|\mathcal{D}_b|),N}, z^{\txt{U}}_{\mathcal{U}_b(1),1}\ldots, z^{\txt{U}}_{\mathcal{U}_b(|\mathcal{U}_b|),N}]$. The superscript $[r]$ denotes the value of the scripted variable at the $r$th iteration.
\begin{IEEEproof}
The proof is based on the description given in \cite[Sec.~III]{Yadav-Dobre-Ansari-access-2017}
\end{IEEEproof}

Using Proposition~\ref{prop:Approx_prop}, \eqref{eq:relax_Problem1} can be solved in a centralized fashion \cite{Yadav-Dobre-Ansari-access-2017} at the cost of heavy information exchange.

\begin{figure}[h]
    \centering
    \includegraphics[width=\columnwidth ]{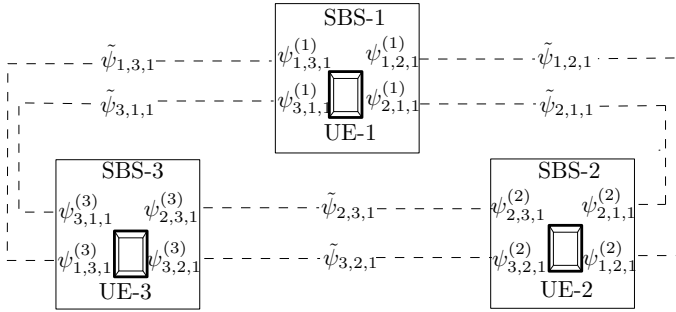}
    \caption{Three small cells network each with one \ac{DL} \ac{UE}. $\mathcal{B}=\{1,2,3\}$, $\bar{\mathcal{D}}_1=\{2,3\}$, $\bar{\mathcal{D}}_2=\{1,3\}$, $\bar{\mathcal{D}}_3\{1,2\}$.}
\label{fig:3SBS_ADMM}
\end{figure}
\section{Distributed Solution}
Owing to the \ac{FD} communications, twice the amount of information exchange is required as compared to the \ac{HD} counterpart for solving the problem in a centralized manner. Furthermore, for a dense network, information exchange requires extra resources that decrease the spectral efficiency of the network. Hence, turning to a distributed approach, where each \ac{SBS} independently designs the beamformers and power allocations locally with minimal information exchange with the rest of the \acp{SBS}, is a necessity.

In order to implement a distributed approach, we take advantage of the separability of the objective function \ac{w.r.t.} each \ac{BS}, and hence \eqref{eq:approx_problem3} can be written equivalently as
\begin{IEEEeqnarray*}{rCl}\label{eq:centralized_approx_problem3}
&\underset{\Xi}{\min} \quad & \Big\{\sum_{b\in \mathcal{B}}\norm{\tilde{\mbf{q}}_{\txt{D},b}}_2 + \sum_{b\in \mathcal{B}} \norm{\tilde{\mbf{q}}_{\txt{U},b}}_2 \mid \eqref{eq:approx_problem3_constr1}- \eqref{eq:approx_problem3_constr9}\Big\},\IEEEyesnumber \quad\,
\end{IEEEeqnarray*}
where $\tilde{\mbf{q}}_{\txt{D},b}$ and 
$\tilde{\mbf{q}}_{\txt{U},b}$ denote the queue deviations of the \ac{DL} and \ac{UL} \acp{UE} associated with $b$, respectively. Observe that the constraints in \eqref{eq:centralized_approx_problem3} are not separable; in particular, constraints  \eqref{eq:approx_problem3_constr2} and \eqref{eq:approx_problem3_constr6} are coupled through the inter-cell \ac{CCI} terms. To this end, we rewrite \eqref{eq:centralized_approx_problem3} as
\begin{IEEEeqnarray*}{lCl}\label{eq:centralized_approx_problem4}
&\displaystyle\underset{\Xi}{\text{min}} \, & \sum_{b\in \mathcal{B}}\norm{\tilde{\mbf{q}}_{\txt{D},b}}_2 + \sum_{b\in \mathcal{B}} \norm{\tilde{\mbf{q}}_{\txt{U},b}}_2 \qquad \IEEEyesnumber
\IEEEyessubnumber* \label{eq:centralized__problem4_obj}\\
&\text{s.t.}& \sigma_n^2 + \sum_{k\in \mathcal{D}_b \setminus \{i\}}\mbf{h}^H_{b_{k},i,n}\mbf{U}_{k,n}\mbf{h}_{b_{k},i,n} + \sum_{\bar{b}\in \bar{\mathcal{B}}_b}\psi^{(b)}_{\bar{b},i,n} \IEEEnonumber \\
&&+ \sum_{j \in \mathcal{U}_b}p_{j,n}|g_{j,i,n}|^2+ \sum_{\bar{b}\in \bar{\mathcal{B}}_b}\phi^{(b)}_{\bar{b},i,n}\leq  \beta_{i,n}\,  \forall i\in \mathcal{D}, \forall n, \qquad\, \label{eq:centralized_approx_problem4_constr1}\\
&& \psi^{(b)}_{b,i,n}\geq \sum_{k\in\mathcal{D}_b}\mbf{h}^H_{b,i,n}\mbf{U}_{k,n}\mbf{h}_{b,i,n} \hfill \forall b,\forall i\in \bar{\mathcal{D}}_b,\forall n,\qquad\,\label{eq:centralized_approx_problem4_constr2} \\
&& \phi^{(b)}_{b,i,n}\geq \sum_{l\in\mathcal{U}_b}p_{l,n}|g_{l,i,n}|^2 \hfill \forall b, \forall i\in\bar{\mathcal{D}}_b, \forall n, \qquad\, \label{eq:centralized_approx_problem4_constr3} \\
&& \bm{\Psi}^{(b)}_{b,j,n}\succeq \sum_{l\in\mathcal{U}_b}p_{l,n}\mbf{h}_{b_{j},l,n}\mbf{h}^H_{b_{j},l,n} \hfill \forall b,\forall j\in\bar{\mathcal{U}}_b,\forall n,\qquad\,\label{eq:centralized_approx_problem4_constr4} \\
&& \bm{\Phi}^{(b)}_{b,j,n}\succeq \sum_{i\in \mathcal{D}_b}\mbf{H}_{b, b_{j},n}\mbf{U}_{i,n}\mbf{H}^{H}_{b, b_{j},n}\hfill \forall b,\forall j\in\bar{\mathcal{U}}_b,\forall n, \qquad\, \label{eq:centralized_approx_problem4_constr5} \\
&& \psi^{(b)}_{b,i,n} = \tilde{\psi}_{b,i,n}\hfill \forall b,\forall i\in \bar{\mathcal{D}}_b,\forall n,\qquad\, \label{eq:centralized_approx_problem4_constr6} \\
&& \psi^{(b)}_{\bar{b},i,n} = \tilde{\psi}_{\bar{b},i,n}\hfill \forall b, \forall \bar{b}\in \bar{\mathcal{B}}_{b},\forall i\in \mathcal{D}_b,\forall n,\qquad\, \label{eq:centralized_approx_problem4_constr61} \\
&& \phi^{(b)}_{b,i,n} = \tilde{\phi}_{b,i,n}\hfill \forall b,\forall i\in \bar{\mathcal{D}}_b,\forall n,\qquad\, \label{eq:centralized_approx_problem4_constr7} \\
&& \phi^{(b)}_{\bar{b},i,n} = \tilde{\phi}_{\bar{b},i,n}\hfill \forall b, \forall \bar{b} \in \bar{\mathcal{B}}_{b},\forall i \in \mathcal{D}_b,\forall n,\qquad\, \label{eq:centralized_approx_problem4_constr71} \\
&& \bm{\Psi}^{(b)}_{b,j,n} = \tilde{\bm{\Psi}}_{b,j,n} \hfill \forall b, \forall j \in \bar{\mathcal{U}}_b,\forall n,\qquad\,\label{eq:centralized_approx_problem4_constr8} \\
&& \bm{\Psi}^{(b)}_{\bar{b},j,n} = \tilde{\bm{\Psi}}_{\bar{b},j,n} \hfill \forall b, \forall \bar{b} \in \bar{\mathcal{B}}_b,\forall j\in \mathcal{U}_b,\forall n,\qquad\,\label{eq:centralized_approx_problem4_constr81} \\
&& \bm{\Phi}^{(b)}_{b,j,n} = \tilde{\bm{\Phi}}_{b,j,n} \hfill \forall b,\forall j\in \bar{\mathcal{U}}_b,\forall n, \qquad\, \label{eq:centralized_approx_problem4_constr9} \\
&& \bm{\Phi}^{(b)}_{\bar{b},j,n} = \tilde{\bm{\Phi}}_{\bar{b},j,n} \hfill \forall b, \forall \bar{b}\in \bar{\mathcal{B}}_{b},\forall j\in \mathcal{U}_b,\forall n, \qquad\, \label{eq:centralized_approx_problem4_constr91} \\
&&\eqref{eq:approx_problem3_constr1}-\eqref{eq:approx_problem3_constr5},
\eqref{eq:approx_problem3_constr7}-\eqref{eq:approx_problem3_constr9},\,
\label{eq:centralized_approx_problem4_constr10}
\end{IEEEeqnarray*} 
where $\bar{\mathcal{B}}_b$, $\bar{\mathcal{D}}_b$ and $\bar{\mathcal{U}}_b$ denote the sets $\mathcal{B}\setminus \{b\}$, $\mathcal{D}\setminus \{\mathcal{D}_b\}$ and $\mathcal{U}\setminus \{\mathcal{U}_b\}$, respectively. $\psi_{b,i,n}$ and $\phi_{b,i,n}$ are newly introduced auxiliary variables, respectively, representing the inter-cell \ac{CCI} caused by the \ac{DL} and \ac{UL} transmissions of \ac{BS} $b$ to the neighboring cells \ac{DL} \ac{UE} $i\in \bar{\mathcal{D}}_b$. Similarly, $\bm{\Psi}_{b,j,n}$ and $\bm{\Phi}_{b,j,n}$ are newly introduced auxiliary variables, respectively, representing the inter-cell \ac{CCI} covariance matrices caused by the \ac{UL} and \ac{DL} transmissions of the \ac{BS} $b$ to the neighboring cells \ac{UL} \ac{UE} $j\in \bar{\mathcal{U}}_b$. The superscript $(\cdot)$ denotes the local copy of the variable. To simplify the decoupling, equality constraints \eqref{eq:centralized_approx_problem4_constr6}-\eqref{eq:centralized_approx_problem4_constr9} are introduced, where $\tilde{\phi}_{b,i,n}, \tilde{\psi}_{b,i,n}, \tilde{\bm{\Phi}}_{b,i,n}$, and $\tilde{\bm{\Psi}}_{b,i,n} \forall b,\forall i\in \bar{\mathcal{D}}_b,\forall n$ are the global variables. Each global variable links the two local variables of the coupled \acp{BS}. For instance, consider a three \ac{SBS} network scenario, as depicted in Fig.~\ref{fig:3SBS_ADMM}, for $b=1\,$, $\tilde{\phi}_{1,2,1}$ represents the same variables $\phi^{(1)}_{1,2,1}$ and $\phi^{(2)}_{1,2,1}$ corresponding to the \ac{BS} $b=1$ and $b=2$, respectively, and so on for all other coupling variables. The equivalence between \eqref{eq:centralized_approx_problem3} and \eqref{eq:centralized_approx_problem4} is due to the fact that constraints \eqref{eq:centralized_approx_problem4_constr1}-\eqref{eq:centralized_approx_problem4_constr5} hold with equality at optimality. 

Observe that \eqref{eq:centralized_approx_problem4} is in a suitable form to apply distributed optimization. The dual decomposition \cite{Boyd-others-FTML-2011} framework offers distributed implementation; however, is suffers from slow convergence. Here, we prefer to use a fast convergence implementation using the \ac{ADMM} \cite{Boyd-others-FTML-2011} framework. For that, we first write the partial augmented Lagrangian dual of \eqref{eq:centralized_approx_problem4} \ac{w.r.t.} the equality constraints as
\begin{IEEEeqnarray*}{lCl}\label{eq:LagrangianDual}
\mathcal{L}(\Xi, \mathcal{X}, \tilde{\mathcal{X}},\hat{\mathcal{X}}) = \sum_{b\in \mathcal{B}}\norm{\tilde{\mbf{q}}_{\txt{D},b}}_2 + \sum_{b\in \mathcal{B}} \norm{\tilde{\mbf{q}}_{\txt{U},b}}_2 &\\
+ \sum_{b\in \mathcal{B}_{b}}\sum_{n=1}^{N}\Big[\sum_{i\in \bar{\mathcal{D}}_b}\theta_{b,i,n}(\psi^{(b)}_{b,i,n}-\tilde{\psi}_{b,i,n}) + \frac{\rho_1}{2}(\psi^{(b)}_{b,i,n}-\tilde{\psi}_{b,i,n})^2&\\
+ \sum_{\bar{b}\in \bar{\mathcal{B}}_b}\sum_{i\in \mathcal{D}_b}\theta_{\bar{b},i,n}(\psi^{(b)}_{\bar{b},i,n}-\tilde{\psi}_{\bar{b},i,n}) + \frac{\rho_1}{2}(\psi^{(b)}_{\bar{b},i,n}-\tilde{\psi}_{\bar{b},i,n})^2&\\
+\sum_{i\in \bar{\mathcal{D}}_b}\omega_{b,i,n}(\phi^{(b)}_{b,i,n}-\tilde{\phi}_{b,i,n}) + \frac{\rho_2}{2}(\phi^{(b)}_{b,i,n}-\tilde{\phi}_{b,i,n})^2&\\
+ \sum_{\bar{b}\in \bar{\mathcal{B}}_b}\sum_{i\in \mathcal{D}_b}\omega_{\bar{b},i,n}(\phi^{(b)}_{\bar{b},i,n}-\tilde{\phi}_{\bar{b},i,n}) + \frac{\rho_2}{2}(\phi^{(b)}_{\bar{b},i,n}-\tilde{\phi}_{\bar{b},i,n})^2&\\
+ \sum_{j\in \bar{\mathcal{U}}_{b}}\text{tr}(\bm{\Theta}_{b,j,n}(\bm{\Psi}^{(b)}_{b,j,n}-\tilde{\bm{\Psi}}_{b,j,n})) + \frac{\rho_3}{2}||\bm{\Psi}^{(b)}_{b,j,n}-\tilde{\bm{\Psi}}_{b,j,n})||^2_2+&\\
\sum_{\bar{b}\in \bar{\mathcal{B}}_b}\sum_{j\in \mathcal{U}_{b}}\text{tr}(\bm{\Theta}_{\bar{b},j,n}(\bm{\Psi}^{(b)}_{\bar{b},j,n}-\tilde{\bm{\Psi}}_{\bar{b},j,n})) + \frac{\rho_3}{2}||\bm{\Psi}^{(b)}_{\bar{b},j,n}-\tilde{\bm{\Psi}}_{\bar{b},j,n})||^2_2 &\\
+ \sum_{j\in \bar{\mathcal{U}}_{b}}\text{tr}(\bm{\Omega}_{b,j,n}(\bm{\Phi}^{(b)}_{b,j,n}-\tilde{\bm{\Phi}}_{b,j,n})) + \frac{\rho_4}{2}||\bm{\Phi}^{(b)}_{b,j,n}-\tilde{\bm{\Phi}}_{b,j,n})||^2_2+&\\
\sum_{\bar{b}\in \bar{\mathcal{B}}_b}\sum_{j\in \mathcal{U}_{b}}\text{tr}(\bm{\Omega}_{\bar{b},j,n}(\bm{\Phi}^{(b)}_{\bar{b},j,n}-\tilde{\bm{\Phi}}_{\bar{b},j,n})) + \frac{\rho_4}{2}||\bm{\Phi}^{(b)}_{\bar{b},j,n}-\tilde{\bm{\Phi}}_{\bar{b},j,n})||^2_2 \Big]\IEEEnonumber &\\
&&\IEEEyesnumber 
\end{IEEEeqnarray*}
where $\rho_1$, $\rho_2$, $\rho_3$, and $\rho_4$ are the positive penalty parameters that controls the rate of convergence. All the local variables are collected into $\mathcal{X}=\{\mathcal{X}_1, \ldots, \mathcal{X}_B\}$, where $\mathcal{X}_b$ collects $\{\phi^{(b)}, \psi^{(b)}, \bm{\Psi}^{(b)}, \bm{\Phi}^{(b)}\}$ and $\phi^{(b)}$ collects $\{\phi^{(b)}_{b,\bar{\mathcal{D}}_b(1),1}, \ldots, \phi^{(b)}_{b,\bar{\mathcal{D}}_b(|\bar{\mathcal{D}}_b|),N},\phi^{(b)}_{\bar{\mathcal{B}}_b(1),\mathcal{D}_b(1),1},\ldots, \allowbreak \phi^{(b)}_{\bar{\mathcal{B}}_b(|\bar{\mathcal{B}}_b|),\mathcal{D}_b(|\mathcal{D}_b|),N} \}$ and similarly $\psi^{(b)}, \bm{\Psi}^{(b)}$, and $\bm{\Phi}^{(b)}$ are represented. Similarly, all the global variables are collected into $\tilde{\mathcal{X}}=\{\tilde{\mathcal{X}}_1, \ldots, \tilde{\mathcal{X}}_B\}$, where $\tilde{\mathcal{X}}_b$ collects $\{\tilde{\phi}_b, \tilde{\psi}_b, \tilde{\bm{\Psi}}_b, \tilde{\bm{\Phi}}_b\}$ and $\tilde{\phi}_b$ collects $\{\tilde{\phi}_{b,\bar{\mathcal{D}}_b(1),1}, \ldots, \tilde{\phi}_{b,\bar{\mathcal{D}}_b(|\bar{\mathcal{D}}_b|),N},\tilde{\phi}_{\bar{\mathcal{B}}_b(1),\mathcal{D}_b(1),1},\ldots, \allowbreak \tilde{\phi}_{\bar{\mathcal{B}}_b(|\bar{\mathcal{B}}_b|),\mathcal{D}_b(|\mathcal{D}_b|),N} \}$ and similarly $\tilde{\psi}, \tilde{\bm{\Psi}}$, and $\tilde{\bm{\Phi}}$ are represented. Similarly, the Lagrangian multipliers are collected in $\hat{\mathcal{X}}=\{\hat{\mathcal{X}}_1\ldots,\hat{\mathcal{X}}_B\}$, where $\hat{\mathcal{X}}_b$ collects $\{\theta_b, \omega_b, \bm{\Theta}_b, \bm{\Omega}_b\}$ with its elements represented similarly as of the local and global variables.

Now, the independent $b$th sub-problem for the $v$th iteration is expressed as
\begin{IEEEeqnarray*}{lCl}\label{eq:sub_problem5}
&{\text{min}} \, & f_b(\Xi_b, \mathcal{X}_b, \tilde{\mathcal{X}}^{[v]}_b,\hat{\mathcal{X}}^{[v]}_b) \IEEEyesnumber 
\IEEEyessubnumber* \label{eq:sub_problem5_obj}\\
&\text{s.t.}& \sigma_n^2 + \sum_{k\in \mathcal{D}_b \setminus \{i\}}\mbf{h}^H_{b_{k},i,n}\mbf{U}_{k,n}\mbf{h}_{b_{k},i,n} + \sum_{\bar{b}\in \bar{\mathcal{B}}_b}\psi_{\bar{b},i,n}^{(b)}\\
&& + \sum_{j \in \mathcal{U}_b}p_{j,n}|g_{j,i,n}|^2  + \sum_{\bar{b}\in \bar{\mathcal{B}}_b}\phi_{\bar{b},i,n}^{(b)}\leq  \beta_{i,n}\,  \forall i \in \mathcal{D}_b, \forall n,\qquad \IEEEyessubnumber* \label{eq:sub_problem5_constr1}\\
&& \psi_{b,i,n}^{(b)}\geq \sum_{k\in\mathcal{D}_b}\mbf{h}^H_{b,i,n}\mbf{U}_{k,n}\mbf{h}_{b,i,n}\hfill \forall i\in \bar{\mathcal{D}}_b,\qquad \, \label{eq:sub_problem5_constr2} 
\end{IEEEeqnarray*}
\begin{IEEEeqnarray*}{lCl}
&& \phi_{b,i,n}^{(b)}\geq \sum_{l\in\mathcal{U}_b}p_{l,n}|g_{l,i,n}|^2 \hfill \forall i\in\bar{\mathcal{D}}_b, \label{eq:sub_problem5_constr3} \IEEEyessubnumber* \\
&& \bm{\Psi}_{b,j,n}^{(b)}\succeq \sum_{l\in\mathcal{U}_b}p_{l,n}\mbf{h}_{b_j,l,n}\mbf{h}^H_{b_j,l,n} \hfill \forall j\in\bar{\mathcal{U}}_b,\label{eq:sub_problem5_constr4} \\
&& \bm{\Phi}_{b,j,n}^{(b)}\succeq\sum_{k\in \mathcal{D}_b}\mbf{H}_{b, b_{j},n}\mbf{U}_{k,n}\mbf{H}^{H}_{b, b_{j},n}\hfill \forall j\in\bar{\mathcal{U}}_b, \IEEEyessubnumber*  \label{eq:sub_problem5_constr5} \\
&& \eqref{eq:centralized_approx_problem4_constr10},
\label{eq:sub_problem5_constr6}
\end{IEEEeqnarray*} 
\begin{IEEEeqnarray*}{lCl}\label{eq:sub_function}
\text{where}\,f_b(\Xi_b, \mathcal{X}_b, \tilde{\mathcal{X}}^{[v]}_b,\hat{\mathcal{X}}^{[v]}_b) = \norm{\tilde{\mbf{q}}_{\txt{D},b}}_2 + \norm{\tilde{\mbf{q}}_{\txt{U},b}}_2 & \\
+ \sum_{n=1}^{N}\Big[\sum_{i\in \bar{\mathcal{D}}_b}\theta_{b,i,n}^{[v]}(\psi^{(b)}_{b,i,n}-\tilde{\psi}_{b,i,n}^{[v]}) + \frac{\rho_1}{2}(\psi^{(b)}_{b,i,n}-\tilde{\psi}_{b,i,n}^{[v]})^2 &\\
+ \sum_{\bar{b}\in \bar{\mathcal{B}}_b}\sum_{i\in \mathcal{D}_b}\theta_{\bar{b},i,n}^{[v]}(\psi^{(b)}_{\bar{b},i,n}-\tilde{\psi}_{\bar{b},i,n}^{v}) + \frac{\rho_1}{2}(\psi^{(b)}_{\bar{b},i,n}-\tilde{\psi}_{\bar{b},i,n}^{[v]})^2 &\\
+\sum_{i\in \bar{\mathcal{D}}_b}\omega_{b,i,n}^{[v]}(\phi^{(b)}_{b,i,n}-\tilde{\phi}_{b,i,n}^{[v]}) + \frac{\rho_2}{2}(\phi^{(b)}_{b,i,n}-\tilde{\phi}_{b,i,n}^{[v]})^2 &\\
+ \sum_{\bar{b}\in \bar{\mathcal{B}}_b}\sum_{i\in \mathcal{D}_b}\omega_{\bar{b},i,n}^{[v]}(\phi^{(b)}_{\bar{b},i,n}-\tilde{\phi}_{\bar{b},i,n}^{[v]}) + \frac{\rho_2}{2}(\phi^{(b)}_{\bar{b},i,n}-\tilde{\phi}_{\bar{b},i,n}^{[v]})^2 &\\
+ \sum_{j\in \bar{\mathcal{U}}_{b}}\text{tr}(\bm{\Theta}_{b,j,n}^{[v]}(\bm{\Psi}^{(b)}_{b,j,n}-\tilde{\bm{\Psi}}_{b,j,n}^{[v]})) + \frac{\rho_3}{2}||\bm{\Psi}^{(b)}_{b,j,n}-\tilde{\bm{\Psi}}_{b,j,n}^{[v]})||^2_2+ &\\
\sum_{\bar{b}\in \bar{\mathcal{B}}_b}\sum_{j\in \mathcal{U}_{b}}\text{tr}(\bm{\Theta}_{\bar{b},j,n}^{[v]}(\bm{\Psi}^{(b)}_{\bar{b},j,n}-\tilde{\bm{\Psi}}_{\bar{b},j,n}^{[v]})) + \frac{\rho_3}{2}||\bm{\Psi}^{(b)}_{\bar{b},j,n}-\tilde{\bm{\Psi}}_{\bar{b},j,n}^{[v]})||^2_2 & \\
+ \sum_{j\in \bar{\mathcal{U}}_{b}}\text{tr}(\bm{\Omega}_{b,j,n}^{[v]}(\bm{\Phi}^{(b)}_{b,j,n}-\tilde{\bm{\Phi}}^{[v]}_{b,j,n})) + \frac{\rho_4}{2}||\bm{\Phi}^{(b)}_{b,j,n}-\tilde{\bm{\Phi}}_{b,j,n}^{[v]})||^2_2+ &\\
\sum_{\bar{b}\in \bar{\mathcal{B}}_b}\sum_{j\in \mathcal{U}_{b}}\text{tr}(\bm{\Omega}_{\bar{b},j,n}^{[v]}(\bm{\Phi}^{(b)}_{\bar{b},j,n}-\tilde{\bm{\Phi}}_{\bar{b},j,n}^{[v]})) + \cfrac{\rho_4}{2}||\bm{\Phi}^{(b)}_{\bar{b},j,n}-\tilde{\bm{\Phi}}_{\bar{b},j,n}^{[v]})||^2_2 \Big],&\,\\ \IEEEnonumber
\end{IEEEeqnarray*}   
and $\hat{\mathcal{X}}^{[v]}_b$ and $\tilde{\mathcal{X}}^{[v]}_b$ denote the collection of fixed Lagrangian multipliers and interference variables updated from the previous iterations. The optimization variables of the problem are $\Xi_b$. After solving \eqref{eq:sub_problem5} for $\Xi_b$, $\bm{\psi}^{(b)}$, $\bm{\phi}^{(b)}$, $\bm{\Psi}^{(b)}$, and $\bm{\Phi}^{(b)}$ $\forall b$ in the $v$th iteration, in the next step, the interference terms are exchanged between \acp{BS} $b$ and $b_i$ as
\begin{IEEEeqnarray}{lCr}   
\tilde{\psi}_{b,i,n}^{[v+1]} = 0.5(\psi_{b,i,n}^{(b)}+\psi_{b,i,n}^{(b_i)}) & \qquad\quad \forall b,\forall i\in \bar{\mathcal{D}}_b, \forall n,\label{eq:GlobalVariablesUpdate1}\\
\tilde{\phi}_{b,i,n}^{[v+1]} = 0.5(\phi_{b,i,n}^{(b)}+\phi_{b,i,n}^{(b_i)}) & \qquad\quad \forall b,\forall i\in \bar{\mathcal{D}}_b, \forall n,\label{eq:GlobalVariablesUpdate2}\\
\tilde{\bm{\Psi}}_{b,j,n}^{[v+1]} = 0.5(\bm{\Psi}_{b,j,n}^{(b)}+\bm{\Psi}_{b,j,n}^{(b_j)}) & \qquad\quad \forall b,\forall j\in \bar{\mathcal{U}}_b, \forall n,\label{eq:GlobalVariablesUpdate3}\\
\tilde{\bm{\Phi}}_{b,j,n}^{[v+1]} = 0.5(\bm{\Phi}_{b,j,n}^{(b)}+\bm{\Phi}_{b,j,n}^{(b_j)}) & \qquad\quad \forall b,\forall j\in \bar{\mathcal{U}}_b, \forall n.\label{eq:GlobalVariablesUpdate4}
\end{IEEEeqnarray}
The final step of the \ac{ADMM} approach is the Lagrangian multipliers update, which is given as 
\begin{IEEEeqnarray*}{lLl}\label{eq:subgradient_1}    
\theta_{b,i,n}^{[v+1]} = [\theta_{b,i,n}^{[v]} + \rho_1^{[v]}(\psi^{(b)}_{b,i,n}-\tilde{\psi}^{[v+1]}_{b,i,n})] \quad \hfill \forall b,\forall i, \forall n, \qquad & \IEEEyesnumber \label{eq:LagrangeMultiplierUpdate1}\\
\omega_{b,i,n}^{[v+1]} = [\omega_{b,i,n}^{[v]} + \rho_2^{[v]}(\phi^{(b)}_{b,i,n}-\tilde{\phi}^{[v+1]}_{b,i,n})]  \quad \hfill \forall b,\forall i, \forall n, \qquad &\IEEEyesnumber \label{eq:LagrangeMultiplierUpdate2}\\
\bm{\Theta}_{b,j,n}^{[v+1]} = [\bm{\Theta}_{b,j,n}^{[v]} + \rho_3^{[v]}(\bm{\Phi}^{(b)}_{b,j,n}-\tilde{\bm{\Phi}}^{[v+1]}_{b,j,n})^{T}] \quad \hfill \forall b, \forall j, \forall n, \qquad &\IEEEyesnumber \label{eq:LagrangeMultiplierUpdate3}\\
\bm{\Omega}_{b,j,n}^{[v+1]} = [\bm{\Omega}_{b,j,n}^{[v]} + \rho_4^{[v]}(\bm{\Psi}^{(b)}_{b,j,n}-\tilde{\bm{\Psi}}^{[v+1]}_{b,j,n})^{T}] \quad \hfill \forall b,\forall j, \forall n. \qquad &\IEEEyesnumber \label{eq:LagrangeMultiplierUpdate4}
\end{IEEEeqnarray*}

Now, in the $r$th \ac{SPCA} iteration index, after the convergence of the \ac{ADMM} procedure, the optimization variables in the set $\Xi$ are updated until the convergence of the \ac{SPCA} procedure.
 The pseudo code of the \ac{ADMM} based distributed algorithm is summarized in Algorithm~\ref{ALG:ADMM_Algo}.

\small
\renewcommand{\baselinestretch}{0.95}{
\begin{algorithm}[H]
\caption{ADMM based distributed iterative algorithm}
\label{ALG:ADMM_Algo}
\begin{algorithmic}[1]
\Require $\mbf{h}$, $\mbf{g}$, $\sigma_n$, $P^{\text{cir}}_{b}$, $P_{b,\text{max}}$, $P_{b}$, $P_{u,\text{max}}$, $\alpha$, $I_{\text{max},1}$, $I_{\text{max},2}$.
\Ensure $\mbf{U}$, $\mbf{p}$.
\State Initialize $r:=0$; $v:=0$, $\bm{\Xi}^{[0]}$, $\tilde{\mathcal{X}}^{[0]}$, and $\hat{\mathcal{X}}^{[0]}=0$;
\Repeat
\Repeat
    \State \parbox[t]{\dimexpr\linewidth-\algorithmicindent}{Solve \eqref{eq:sub_problem5} for $\bm{\Xi}_b^{[r]}, \mathcal{X}_b^{[v]},\tilde{\mathcal{X}}_b^{[v]}\,\forall b\in \mathcal{B}$ using $\hat{\mathcal{X}}_b^{[v]}$}
    \State \parbox[t]{\dimexpr\linewidth-\algorithmicindent}{Exchange $\mathcal{X}_b^{[v]}$ among \acp{BS}}
    \State \parbox[t]{\dimexpr\linewidth-\algorithmicindent}{Update $\tilde{\mathcal{X}}_b^{[v+1]}$ using  \eqref{eq:GlobalVariablesUpdate1} -- \eqref{eq:GlobalVariablesUpdate4}}
    \State \parbox[t]{\dimexpr\linewidth-\algorithmicindent}{Update $\hat{\mathcal{X}}_b^{[v+1]}$ using  \eqref{eq:LagrangeMultiplierUpdate1} -- \eqref{eq:LagrangeMultiplierUpdate4}}    
        \State Set $v:=v+1$
\Until \parbox[t]{\dimexpr\linewidth-\algorithmicindent}{{Convergence of \ac{ADMM} algo. or $v\geq I_{\text{max},2}$}}
\State \parbox[t]{\dimexpr\linewidth-\algorithmicindent}{Update $\bm{\Xi}^{[r+1]}=\bm{\Xi}^{\star}$;}
\State $r:= r+1$; $v:=0$
\Until{Queue convergence or $r\geq I_{\text{max},1}$}
\State Perform randomization to extract a rank-one solution
\end{algorithmic}
\end{algorithm}
\normalsize

\begin{table}[h!]
\renewcommand{\arraystretch}{1}
\caption{Simulation Parameters}
\label{tab:simulation_parameters}
\centering
\begin{tabular}{c|c}
\hline
\bfseries Parameters & \bfseries Value\\
\hline
\hline
No. of antennas & $M_T=2$, $M_R=2$ \\
\hline
No. of sub-carriers & $N=2$ \\
\hline
Cell radius  & \ac{MBS}: $500$ m, \ac{SBS}: $50$ m \\
\hline
Maximum transmit power & \ac{SBS}: $24$ dBm, \ac{UE}: $23$ dBm\\
\hline
Circuit power & $30$ dBm\\
\hline
Bandwidth & $10$ MHz\\
\hline
Intensity & \ac{SBS}: $\lambda_s=10$, \ac{UE}: $\lambda_u=2\lambda_s$, \\
\hline
Thermal noise density and SI & $-174$ dBm/Hz, $\sigma^2_{\text{SI}}=-110$ dB\\
\hline
DE parameter & $\alpha = 0.1$ \\
\hline
Noise figure & \ac{SBS}: $13$ dB, \ac{UE}: $9$ dB \\
\hline
Path loss (in dB) \ac{SBS}-to-\ac{SBS}	& LOS: $98.4+20.9\log_{10}(d)$\\
where $d$ is in km	& NLOS: $169.36+40\log_{10}(d)$ \\
\hline
Path loss (in dB) \ac{UE}-to-\ac{SBS}	& LOS: $103.8+20.9\log_{10}(d)$\\
where $d$ is in km	& NLOS: $145.4+37.5\log_{10}(d)$ \\
\hline
Path loss (in dB) \ac{UE}-to-\ac{UE}	& LOS: $98.5+20\log_{10}(d)$\\
where $d$ is in km	& NLOS: $175.78+40\log_{10}(d)$ \\\hline
\end{tabular}
\end{table}
\section{Numerical Results and Discussions}
The numerical simulation results obtained by using the distributed Algorithm~\ref{ALG:ADMM_Algo} are presented in this section. A typical outdoor deployment scenario with a circular macro-cell area in the plane $\mathbb{R}^2$ is considered. One \ac{MBS} located at the origin and ten randomly deployed \acp{SBS}, i.e., $B=10$, whose locations follow an independent \ac{PPP} $\Phi_s \in \mathbb{R}^2$ with intensity $\lambda_s$, are considered. We assume a total of two \ac{DL} and two \ac{UL} \acp{UE} within each \ac{SBS} and they are randomly located according to the \ac{PPP} $\Phi_u\in \mathbb{R}^2$ with intensity $\lambda_u$. Hence, the total number of \acp{UE} in the network is $K_{\txt{D}}=K_{\txt{U}}= 20$. The maximum transmission powers of \acp{SBS} and \acp{UE} are fixed and given by $P_{b,\text{max}}$ and $P_{\text{max}}$, respectively. The Rician fading model is considered to model the \ac{SI} channel between the co-located transmitter-receiver antenna pair of an \ac{SBS} with distribution $\mathcal{CN}(\sqrt{\sigma^2_{\text{SI}}K/(1+K)}\mbf{H}_{\text{SI}},(\sigma^2_{\text{SI}}/(1+K))\mbf{I}_{M_R}\otimes \mbf{I}_{M_T})$, where $\mbf{H}_{\text{SI}}$ is a deterministic matrix and  $K$ is the Rician factor with value $1$, and $\sigma^2_{\text{SI}}$ is the \ac{SI} variance. The rest of the channels in the system are assumed to be Rayleigh faded and the effect of the path and shadowing loss is already included in them. All other simulation parameters used are listed in Table~\ref{tab:simulation_parameters}. We especially consider three system scenarios for comparison, which are referred to as: i) Setup-A: \acp{SBS} are powered by the grid source; ii) Setup-B: \acp{SBS} are powered by a renewable energy source; and iii) Setup-C: \acp{SBS} are powered by a renewable energy source and consume energy for decoding \ac{UL} \acp{UE} data. The number of bits waiting in the data buffer of each \ac{DL} and \ac{UL} \ac{UE} are stored in vectors $Q^{\txt{D}}=[6\,     7\,     4\,     5\,     3\,     2\,     2\,     2\,     2\,     3\,     1\,     1\,     2\,     2\,     2\,     3\,     2\,     2\,     3\,     7]$ and $Q^{\txt{U}}=[3\,     7\,     3\,     5\,     7\,     3\,     2\,     3\,     1\,     3\,     3\,     3\,     3\,     1\,     2\,     2\,     2\,     2\,     3\,     2\,     1\,     1]$, respectively. 

\begin{figure}[h!]
\centerline{\epsfig{figure=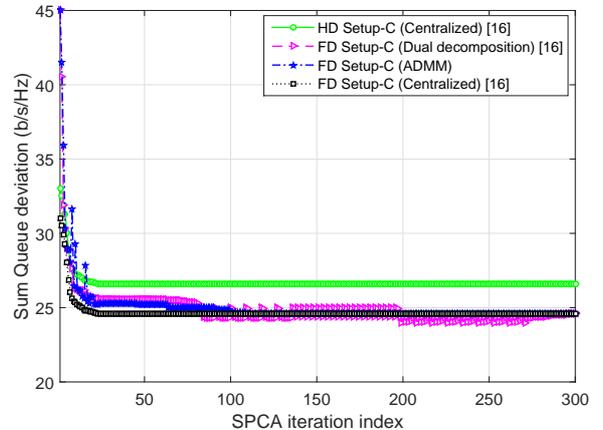,width=3.0in}}
\begin{center}
\caption{Convergence of the proposed \ac{ADMM}-based \ac{RAOFDS} algorithm with respect to the \ac{SPCA} iteration index.}\label{fig:Convergence_Algorithm-1}
\end{center}
\end{figure}

We first compare the convergence of the proposed \ac{ADMM} based distributed Algorithm~\ref{ALG:ADMM_Algo} with the centralized and dual decomposition based distributed algorithms \cite{Yadav-Dobre-Ansari-access-2017} in Fig.~\ref{fig:Convergence_Algorithm-1}. The figure plots the total number of bits that remain in the network after each \ac{SPCA} iteration step under the system  Setup-C. It can be observed that the centralized algorithm converges faster than both \ac{ADMM} and dual decomposition based distributed algorithms. However, among the distributed algorithms, the \ac{ADMM} approach converges faster by taking approximately 200 iterations lesser than the dual decomposition approach, which takes 300 iterations. Note that all three algorithms converge to the same value of the queue deviation. 

In Fig~\ref{fig:Convergence_Algorithm-1}, the performance of the \ac{FD} and \ac{HD} \acp{SBS} is also compared. As expected, the \ac{FD} \acp{SBS} achieve lower total queue deviation than the conventional \ac{HD} \acp{SBS}. In next two examples, we only consider the performances of the \ac{FD} \acp{SBS} for the presentation clarity.

\begin{figure}[t]
\centerline{\epsfig{figure=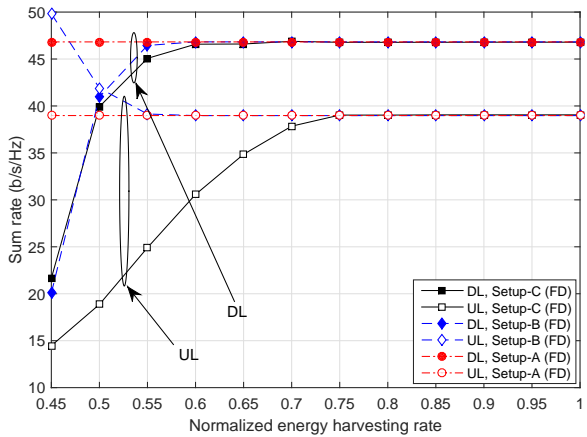,width=3.0in}}
\begin{center}
\caption{\ac{DL} and \ac{UL} sum rate of the network with different normalized \ac{EH} arrival rates at each \ac{SBS}.}\label{fig:Sumrate_vs_EHrate_2N_10SCells_100SI}
\end{center}
\end{figure}
Fig.~\ref{fig:Sumrate_vs_EHrate_2N_10SCells_100SI} shows the sum rate performance achieved by the network with different values of the normalized energy arrival rates, i.e., $P_{b,\text{H}}/(P_b^{\text{cir}}+5P_{b,\text{max}})$ at the \acp{SBS} under the Setup-B and Setup-C. For comparison, the sum rate of Setup-A is plotted; however, it is independent of the energy arrivals. In the low \ac{EH} rate regime, for Setup-B, the sum rate is higher for \ac{UL} as the \ac{SBS} has lower energy availability for the \ac{DL} \acp{UE}; hence, it produces low interference to the \ac{UL} \acp{UE}. On the other hand, the \ac{DL} transmissions achieve higher sum rate in the high \ac{EH} rate regime. Consequently, the \ac{UL} transmissions receive higher interference from the high power \ac{DL} transmissions. This behavior is reversed for Setup-C, where the \ac{DL} sum rates dominate in all \ac{EH} rate regimes over the \ac{UL} sum rates. The reason for this is that, in Setup-C, the \ac{SBS} shares the harvested energy among the \ac{DL} and \ac{UL} \acp{UE}. Therefore, lower energy availability at the \ac{SBS} limits the \ac{UL} \acp{UE} from using lower transmit power that consequently introduces less interference into the \ac{DL} transmissions.   

\begin{figure}[t]
\centerline{\epsfig{figure=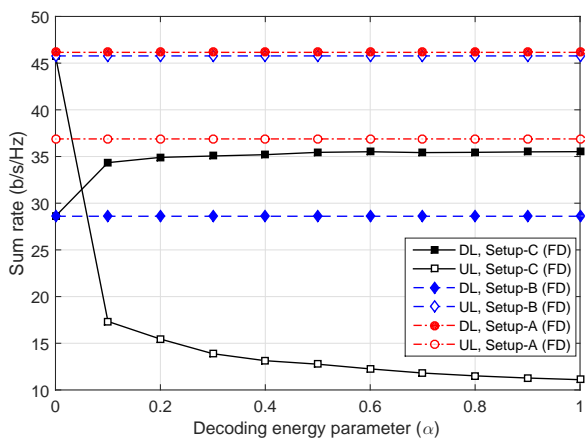,width=3.0in}}
\begin{center}
\caption{\ac{DL} and \ac{UL} sum rate versus the \ac{DE} parameters used by each \ac{SBS}.}\label{fig:Sumrate_vs_DE_2N_10SCells_100SI}
\end{center}
\end{figure}
Fig.~\ref{fig:Sumrate_vs_DE_2N_10SCells_100SI} show the sum rate achieved by the network under Setup-C with different values of the \ac{DE} parameter. For comparison purposes, the figure also plots the sum rates achieved under the Setup-A and -B, which are independent of the \ac{DE} parameter. Observe that the sum rate achieved by the \ac{UL} \acp{UE} decreases with the increase in the portion of \ac{DE} consumed at the \ac{SBS}. This is because the \ac{UL} \acp{UE} rates are now determined by the availability of the \ac{DE} at the \ac{SBS}. For instance, if the value of the \ac{DE} parameter is small, the \ac{SBS} allocates a small portion of the energy for the \ac{UL} \acp{UE} decoding. This essentially means that the \ac{UL} \acp{UE} cannot be decoded if transmitted at higher rate and \ac{UL} \acp{UE} need to transmit with lower power. Consequently, a lower interference is experienced by the \ac{DL} \acp{UE}, and thus, the sum rate improves as compared to Setup-B. High \ac{DE} parameter values further restrict the \ac{UL} \acp{UE} from transmitting at lower power, and hence, \ac{DL} \acp{UE} experience low interference. 

\section{Conclusion}
The performance of densely deployed \ac{FD} small cells is studied at the network level. The \acp{SBS} are dependent on the renewable energy source for its transceiver operations. The \ac{UL} \acp{UE} rate-dependent decoding energy is included in the total energy consumption model at the \acp{SBS}. Hence, the energy harvested at the \ac{SBS} must be optimally shared among the \ac{DL} and \ac{UL} \acp{UE}. A joint beamformer and power allocation design, which minimizes the \acp{UE} data buffer lengths, is proposed. Furthermore, the proposed optimization problem implicitly solves the problem of sub-carrier allocation and \acp{UE} scheduling. A sub-optimal and iterative \ac{SPCA}-based approach is used to circumvent the non-convex nature of the problem. A fast-convergent algorithm based on the \ac{ADMM} framework is proposed to solve the optimization problem distributively. Simulations are used to compare the performances of the proposed design under the practical energy consumption and casualty constraints with the case when the \ac{DE} is not considered. Results show the performance gap and advocate the need for redesigning the beamformers and power allocations.
          
\renewcommand{\baselinestretch}{1}
\bibliographystyle{IEEEtran}
\bibliography{IEEEabrv,conf_short,jour_short,Ref_SmallCellFullDuplex}

\begin{thebibliography}{10}
\providecommand{\url}[1]{#1}
\csname url@samestyle\endcsname
\providecommand{\newblock}{\relax}
\providecommand{\bibinfo}[2]{#2}
\providecommand{\BIBentrySTDinterwordspacing}{\spaceskip=0pt\relax}
\providecommand{\BIBentryALTinterwordstretchfactor}{4}
\providecommand{\BIBentryALTinterwordspacing}{\spaceskip=\fontdimen2\font plus
\BIBentryALTinterwordstretchfactor\fontdimen3\font minus
  \fontdimen4\font\relax}
\providecommand{\BIBforeignlanguage}[2]{{%
\expandafter\ifx\csname l@#1\endcsname\relax
\typeout{** WARNING: IEEEtran.bst: No hyphenation pattern has been}%
\typeout{** loaded for the language `#1'. Using the pattern for}%
\typeout{** the default language instead.}%
\else
\language=\csname l@#1\endcsname
\fi
#2}}
\providecommand{\BIBdecl}{\relax}
\BIBdecl

\bibitem{Bhushan-Li-Malladi-Gilmore-Brenner-Damnjanovic-Sukhavasi-Patel-Geirhofer-mcom-15}
N.~Bhushan \emph{et~al.}, ``Network densification: {T}he dominant theme for
  wireless evolution into 5{G},'' \emph{{IEEE} Commun. Mag.}, vol.~52, no.~2,
  pp. 82--89, Feb. 2014.

\bibitem{Huang-Han-Ansari-CST-2015}
X.~Huang, T.~Han, and N.~Ansari, ``On green-energy-powered cognitive radio
  networks,'' \emph{{IEEE} Commun. Surveys Tut.}, vol.~17, no.~2, pp. 827--842,
  2nd Quart. 2015.

\bibitem{he2014recursive}
P.~He, L.~Zhao, S.~Zhou, and Z.~Niu, ``Recursive waterfilling for wireless
  links with energy harvesting transmitters,'' \emph{{IEEE} Trans. Veh.
  Technol.}, vol.~63, no.~3, pp. 1232--1241, Mar. 2014.

\bibitem{Ahmed-Ikhlef-Ng-Schober-twcom-13}
I.~Ahmed, A.~Ikhlef, D.~W.~K. Ng, and R.~Schober, ``Power allocation for an
  energy harvesting transmitter with hybrid energy sources,'' \emph{{IEEE}
  Trans. Wireless Commun.}, vol.~12, no.~12, pp. 6255--6267, Dec. 2013.

\bibitem{Arafa-Ulkus-jsac-2015}
A.~Arafa and S.~Ulukus, ``Optimal policies for wireless networks with energy
  harvesting transmitters and receivers: {E}ffects of decoding costs,''
  \emph{{IEEE} J. Select. Areas Commun.}, vol.~33, no.~12, pp. 2611--2625, Dec.
  2015.

\bibitem{Yadav-Nguyen-Ajib-tcom-2016}
A.~Yadav, T.~M. Nguyen, and W.~Ajib, ``Optimal energy management in hybrid
  energy small cell access points,'' \emph{{IEEE} Trans. Commun.}, vol.~64,
  no.~12, pp. 5334--5348, Dec. 2016.

\bibitem{he2016optimal}
P.~He and L.~Zhao, ``Optimal power allocation for maximum throughput of general
  {MU}-{MIMO} multiple access channels with mixed constraints,'' \emph{{IEEE}
  Trans. Commun.}, vol.~64, no.~3, pp. 1042--1054, Mar. 2016.

\bibitem{he2016non}
------, ``Non-commutative composite water-fillings for energy harvesting and
  smart power grid hybrid system with peak power constraints,'' \emph{{IEEE}
  Trans. Veh. Technol.}, vol.~65, no.~4, pp. 2026--2037, Apr. 2016.

\bibitem{Hong-others-CM-2014}
S.~Hong \emph{et~al.}, ``Applications of self-interference cancellation in 5{G}
  and beyond,'' \emph{{IEEE} Commun. Mag.}, vol.~52, no.~2, pp. 114--121, Feb.
  2014.

\bibitem{Bharadia-Mcmilin-Katti-sigcomm-2013}
D.~Bharadia, E.~McMilin, and S.~Katti, ``Full duplex radios,'' \emph{SIGCOMM
  Comput. Commun. Rev.}, vol.~43, no.~4, pp. 375--386, Aug. 2013.

\bibitem{Nguyen-Tran-Pirinen-Latva-aho-twc-2014}
D.~Nguyen, L.~N. Tran, P.~Pirinen, and M.~Latva-aho, ``On the spectral
  efficiency of full-duplex small cell wireless systems,'' \emph{{IEEE} Trans.
  Wireless Commun.}, vol.~13, no.~9, pp. 4896--4910, Sep. 2014.

\bibitem{Goyal-Liu-Panwar-tvt-2016}
S.~Goyal, P.~Liu, and S.~S. Panwar, ``User selection and power allocation in
  full duplex multi-cell networks,'' \emph{{IEEE} Trans. Veh. Technol.},
  vol.~PP, no.~99, pp. 1--15, Jun. 2016.

\bibitem{Cui-Goldsmith-Bahai-jsac-04}
S.~Cui, A.~Goldsmith, and A.~Bahai, ``Energy-efficiency of {MIMO} and
  cooperative {MIMO} techniques in sensor networks,'' \emph{{IEEE} J. Select.
  Areas Commun.}, vol.~22, no.~6, pp. 1089--1098, Aug. 2004.

\bibitem{Chen-other-jsac-2016}
L.~Chen \emph{et~al.}, ``Green full-duplex self-backhaul and energy harvesting
  small cell networks with massive {MIMO},'' \emph{{IEEE} J. Select. Areas
  Commun.}, vol.~34, no.~12, pp. 3709--3724, Dec. 2016.

\bibitem{Rubio-Pascual-Iserte-twc-2014}
J.~Rubio and A.~Pascual-Iserte, ``Energy-aware broadcast multiuser-{MIMO}
  precoder design with imperfect channel and battery knowledge,'' \emph{{IEEE}
  Trans. Wireless Commun.}, vol.~13, no.~6, pp. 3137--3152, Jun. 2014.

\bibitem{Yadav-Dobre-Ansari-access-2017}
A.~Yadav, O.~A. Dobre, and N.~Ansari, ``Energy and traffic aware full-duplex
  communications for {5G} systems,'' \emph{{IEEE} Access}, Mar. 2017 (to
  appear).

\bibitem{Boyd-others-FTML-2011}
S.~Boyd \emph{et~al.}, ``Distributed optimization and statistical learning via
  the alternating direction method of multipliers,'' \emph{Found. Trends Mach.
  Learn.}, vol.~3, no.~1, pp. 1--122, 2011.

\bibitem{Beck-Ben-Tal-Tetruashvili-jgo-2010}
A.~Beck, A.~Ben-Tal, and L.~Tetruashvili, ``A sequential parametric convex
  approximation method with applications to nonconvex truss topology design
  problems,'' \emph{J. Global Optim., Springer}, vol.~47, no.~1, pp. 29--51,
  May 2010.

\end{thebibliography}

\end{document}